\documentclass[superscriptaddress,nobibnotes,amsmath,amssymb,notitlepage,twocolumn,pra,reprint,longbibliography]{revtex4-1}

\usepackage[colorlinks=true,linkcolor=blue,citecolor=blue,urlcolor=blue]{hyperref}

\usepackage{setspace} 
\usepackage{graphicx} \graphicspath{{./img/}} 
\usepackage{amsmath}
\usepackage{color}
\usepackage{amsmath}
\usepackage{amssymb}
\usepackage{verbatim}
\usepackage{latexsym}
\usepackage{enumerate} 
\usepackage{bm} 
\usepackage{ulem} 

\newcommand*{\heading}[1]{\vspace{2mm}\noindent\belowpdfbookmark{#1}{#1}{\bfseries{\large #1 }}\ignorespaces\vspace{1mm}}
\let\section\heading 
\newcommand*{\subheading}[1]{\noindent\belowpdfbookmark{#1}{#1}{\bfseries{#1: }}\ignorespaces}
\let\subsection\subheading 

\begin{document}

\title{
Electronic structure of fullerene nanoribbons
}



\newcommand{\TCM}{Theory of Condensed Matter Group, Cavendish Laboratory, University of Cambridge, J.\,J.\,Thomson Avenue, Cambridge CB3 0HE, UK}
\newcommand{\HarvardFAS}{Harvard University, Cambridge, Massachusetts 02138, USA}


 \author{Bo Peng}
 \email{bp432@cam.ac.uk}
 \affiliation{Theory of Condensed Matter Group, Cavendish Laboratory, University of Cambridge, Cambridge CB3 0HE, United Kingdom}

 \author{Michele Pizzochero}
 \email{mp2834@bath.ac.uk}
 \affiliation{Department of Physics, University of Bath, Bath BA2 7AY, United Kingdom}
 \affiliation{School of Engineering and Applied Sciences, Harvard University, Cambridge, Massachusetts 02138, United States}

\date{\today}

\begin{abstract}
Using first-principles calculations, we examine the electronic structure of quasi-one-dimensional fullerene nanoribbons derived from two-dimensional fullerene networks. Depending on the edge geometry and width, these nanoribbons exhibit a rich variety of properties, including direct and indirect band gaps, positive and negative effective masses, as well as dispersive and flat bands. Our findings establish a comprehensive understanding of the electronic properties of fullerene nanoribbons, with potential implications for the design of future nanoscale devices.

\end{abstract}

\maketitle


\section{Introduction}

\noindent The edges of graphene exhibit intriguing physical properties, which have motivated the fabrication of graphene nanoribbons---a class of nanoscale materials composed of quasi-one-dimensional strips of hexagonally bonded carbon atoms. The exploration of graphene nanoribbons has opened  new avenues for both fundamental research \cite{Nakada1996,Yazyev2013} and future technology\,\cite{Chen2020b,Wang2021c}. Their structural and electronic properties can be controlled through width and edge geometry\,\cite{Son2006}, serving as new degrees of freedom to achieve target structures and functionalities, e.g., heterojunctions with tunable band gaps\,\cite{Chen2015b,Nguyen2017,Cernevics2020}. Additionally, the electronic properties of graphene nanoribbons can be controlled via chemical or electric approaches, leading to rich electronic phases, including Dirac semimetallic\,\cite{Raza2008,Pizzochero2021b}, half-metallic\,\cite{Son2006a,Pizzochero2022}, magnetic\,\cite{Yazyev2010, Ma2025}, and topological\,\cite{Groning2018, Rizzo2018, Tepliakov2023} phases.

The family of carbon-based two-dimensional materials has recently expanded with the introduction of monolayer fullerene (C$_{60}$) networks \cite{Hou2022}, which offer a promising platform to realize potential applications in photocatalysis\,\cite{Peng2022c,Jones2023,Shearsby2025,Wu2025,Wang2023}, thermal devices\,\cite{Shaikh2025,Yu2022,Meirzadeh2023}, nanofiltration\,\cite{Tong2023,Chen2024,Tong2024}, and photodetectors\,\cite{Kayley2025,Tromer2022,Zhang2025} within an ultrathin ($< 1$\,nm) molecular nanostructure. Yet, a thorough investigation of the properties on the edges of such monolayers is missing, with earlier studies being restricted only to polymeric C$_{60}$ chain with a width of a single molecule\,\cite{Chauvet1994,Xu1995,Springborg1995,Nunez-Regueiro1995,Marques1996,Stephens1994,Gunnarsson1997,Huq2001,Belosludov2003,Belosludov2006}. Understanding the impact of edges on monolayer fullerene networks  is an issue is of particular relevance to experiments, given  that many structural phases of these networks tend to split into nanoribbons with increasing temperature or under mechanical strain\,\cite{Peng2023,Ribeiro2022,Ying2023}.

Here, we employ first-principles calculations to investigate the structural and electronic properties of fullerene nanoribbons derived from the experimentally known structural phases of monolayer fullerene networks. Depending on the edge geometry and width, we show that a variety of  electronic properties, e.g., direct/indirect band gaps and negative/positive carrier effective masses, can be obtained from the same parent monolayers. While electronic properties for certain nanoribbon structures converge to those of their monolayer counterpart as the two-dimensional limit is approched, other nanoribbons exhibit edge-induced states with distinct properties, such as flat-band features. Our work forms the basis for designing fullerene-based nanostructures with unique advantages such as scalability and controllability.

\section{Methodology} 

\noindent Density functional theory (DFT) calculations were performed using the {\sc siesta} package\,\cite{Soler2002,Artacho2008,Garcia2020} under the spin-polarized, generalized-gradient approximation (GGA) of  Perdew, Burke, and Ernzerhof (PBE) \cite{Perdew1996}. A double-$\zeta$ plus polarization (DZP) basis set was used with an energy cutoff of 400\,Ry and a reciprocal space sampling of 18 $k$-points along the periodic direction. A vacuum spacing in the non-periodic directions larger than 20\,\AA, was used throughout. Both the lattice constant and atomic positions were fully relaxed using the conjugate gradient method\,\cite{Payne1992} with a tolerance on forces of 0.02\,eV/\AA.  

\begin{figure}
\centering
\includegraphics[width=\linewidth]{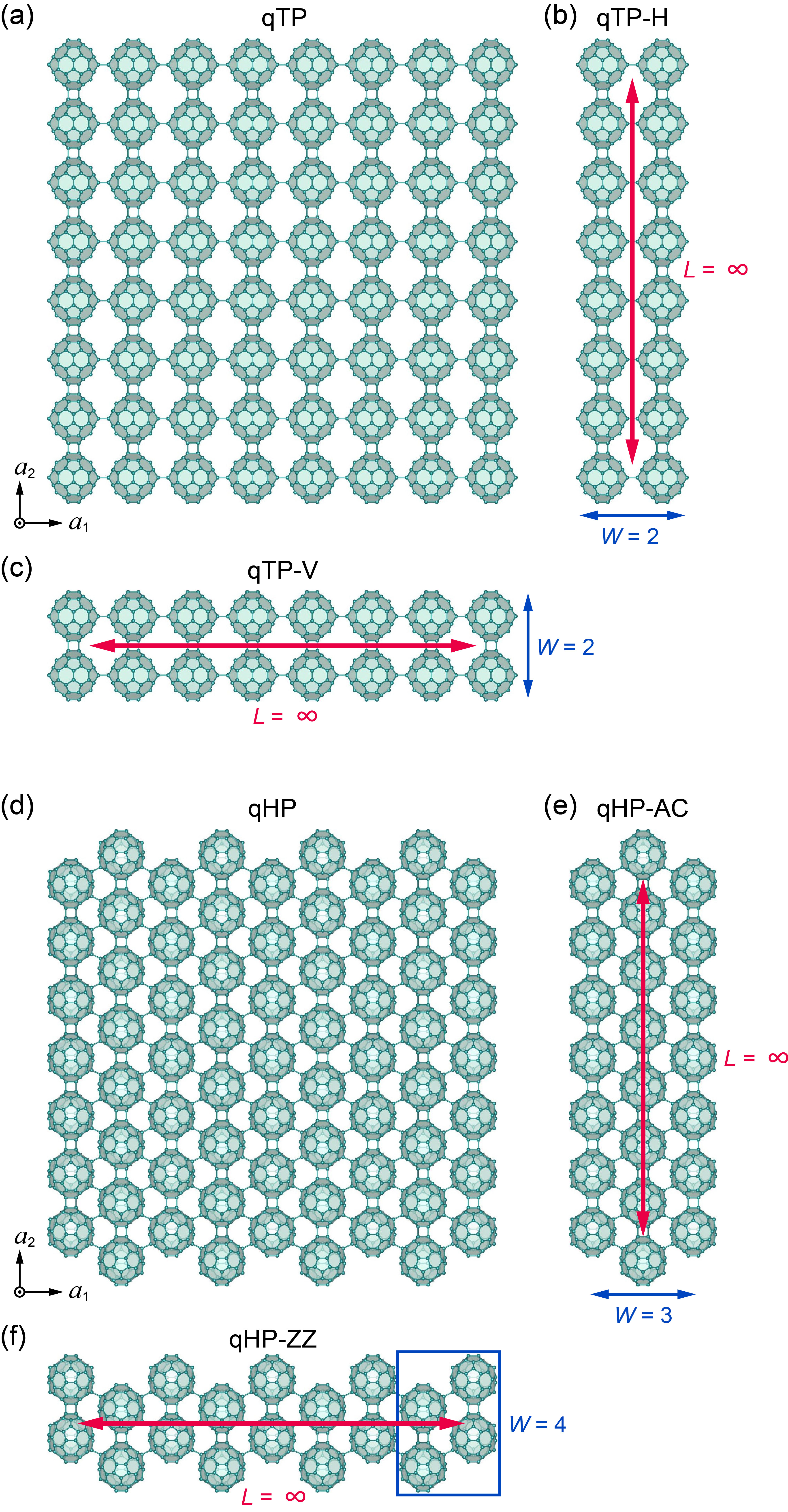}
\caption{
Crystal structures of (a) monolayer qTP fullerene networks, (b) qTP-H and (c) qTP-V nanoribbons, as well as (d) monolayer qHP fullerene networks, (e) qHP-AC and (f) qHP-ZZ nanoribbons.
}
\label{crystals} 
\end{figure}

\section{Results and Discussion} 

\subsection{From 2D to 1D} 
We consider two experimentally known crystalline phases of monolayer fullerene networks, i.e., the quasi-tetragonal phase (qTP) and the quasi-hexagonal phase (qHP)\,\cite{Hou2022}.
Fig.\,\ref{crystals}(a) shows the crystal structures of the qTP phase. In monolayer qTP networks, neighboring carbon cages are linked by vertical [2+2] cycloaddition bonds along the $a_1$ direction, while the C$_{60}$ molecules are connected by horizontal [2\,+\,2] cycloaddition bonds along $a_2$. Therefore, we denote the nanoribbons forming along the $a_2$ direction in Fig.\,\ref{crystals}(b) as qTP-H nanoribbons and the nanoribbons forming along $a_2$ in Fig.\,\ref{crystals}(c) as qTP-V nanoribbons. 

Different from the qTP C$_{60}$ networks, the qHP shown in Fig.\,\ref{crystals}(d) has a closely-packed structure with C$-$C single bonds connecting adjacent fullerene units along the diagonal of $a_1$ and $a_2$ and nearly-horizontal [2\,+\,2] cycloaddition bonds connecting neighboring units along $a_2$. Consequently, the nanoribbons forming along the $a_2$ direction have the armchair structure in Fig.\,\ref{crystals}(e) and we denote them as qHP-AC nanoribbons. The nanoribbons forming along the $a_1$ direction exhibit a zigzag-like structure in Fig.\,\ref{crystals}(f), and we denote them as qHP-ZZ nanoribbons. Overall, there are four edge geometries: qTP-H, qTP-V, qHP-AC, and qHP-ZZ. We quantify the width of the nanoribbons, $W$, as the number of C$_{60}$ molecules across the periodic direction. We study nanoribbon structures with $W>1$, contrary to a purely 1D polymeric fullerene chain reported previously\,\cite{Jones2023}.

\begin{figure*}
\centering
\includegraphics[width=\linewidth]{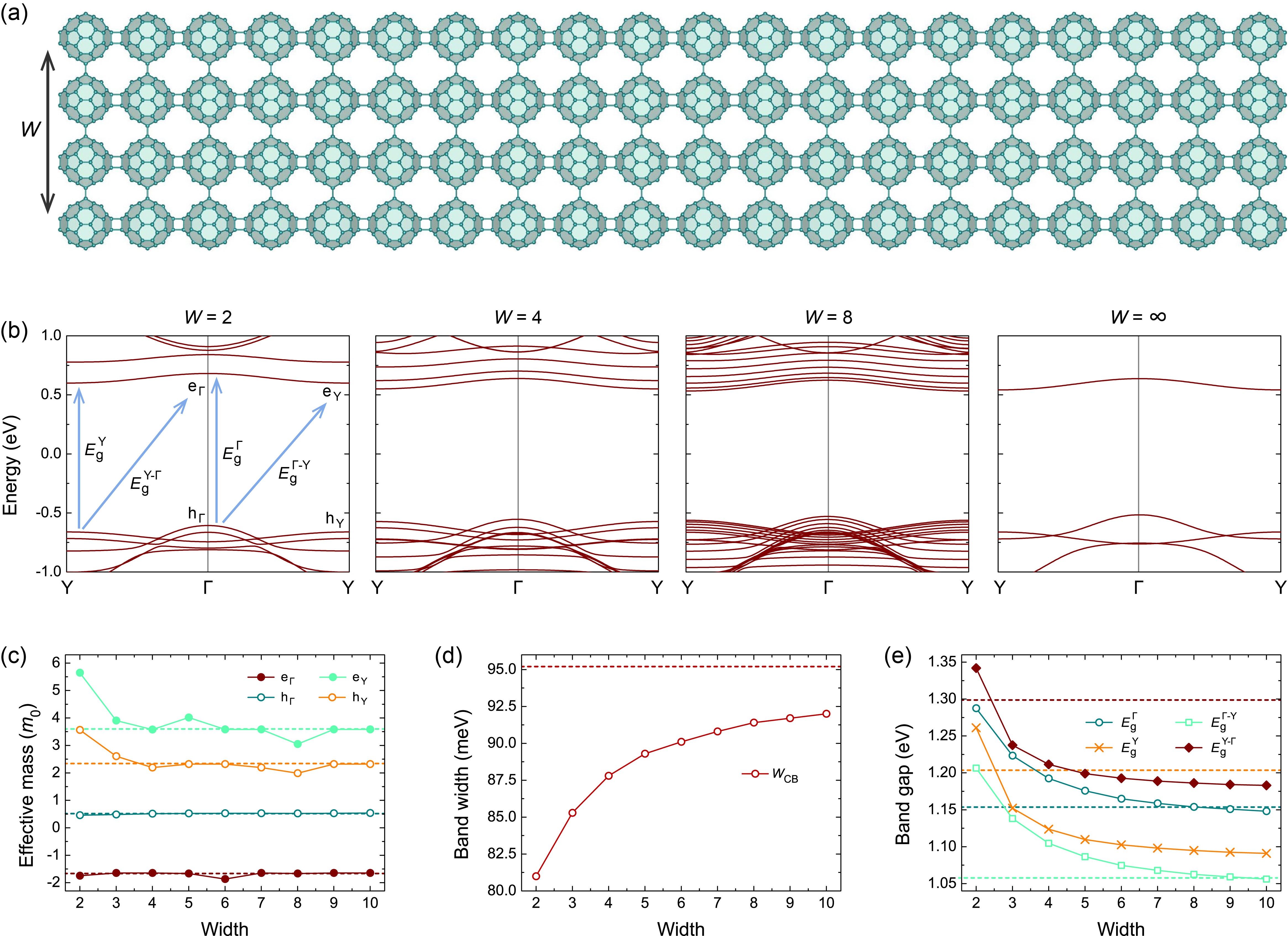}
\caption{
(a) Crystal structures of qTP-H nanoribbons with a representative width  $W$ of 4. (b) Band structures, (c) effective masses, (d) band widths, and (e) band gaps of qTP-H nanoribbons as a function of $W$.
}
\label{qTP-H} 
\end{figure*}

\vspace{1mm}

\subsection{qTP-H} 
We first focus on qTP-H nanoribbons. Fig.\,\ref{qTP-H}(a) shows the representative crystal structure of a qTP-H nanoribbon with $W = 4$. This nanoribbon has a lattice constant of 9.05\,\AA\ and a width of 34.33\,\AA. The space group of qTP-H nanoribbon is $Pmmm$ (No.\,47) with inversion symmetry, as well as $C_2$ rotational and mirror symmetry with respect to the $x$, $y$ and $z$ axes.

In Fig.\,\ref{qTP-H}(b), we show the evolution of the band structures as a function of $W$, where similar features are observed by incrasing $W$ from 2 to $\infty$. Increasing $W$ leads to more replicas of electronic states. As an example, the two lowest conduction bands for $W=2$ show similar curvatures, while the conduction bands with similar curvatures are doubled for $W=4$. There are two valence band maxima (VBM) at $\Gamma$ and Y, denoted as h$_{\Gamma}$ and h$_{\rm Y}$, respectively, as well as one conduction band minimum (CBM) at Y and the negative-curvature, lowest conduction band around $\Gamma$, denoted as e$_{\rm Y}$ and e$_{\Gamma}$, respectively). 

We then investigate the evolution of effective masses of the band edges in Fig.\,\ref{qTP-H}(c). For band edges at Y, the effective masses for h$_{\rm Y}$ and e$_{\rm Y}$ decrease with $W$, until converging to the monolayer masses when $W\geq9$. The band edges at Y have large effective masses with $m({\rm h}_{\rm Y})>2\,m_0$ and $m({\rm e}_{\rm Y})>3\,m_0$, as the band edges around Y are less dispersive than those around $\Gamma$. For h$_{\Gamma}$ and e$_{\Gamma}$, the effective masses remain nearly unchanged and are comparable to the monolayer masses. The band edges in the vicinity of $\Gamma$ have smaller effective masses as they are more dispersive, especially for h$_{\Gamma}$ with effective masses around 0.5\,$m_0$. Interestingly, e$_{\Gamma}$ always has negative effective mass $m({\rm e}_{\Gamma}) \sim -1.7\,m_0$ even for $W=\infty$. Therefore, the electron and hole at $\Gamma$ have negative total mass [$m({\rm e})+m({\rm h})$] but positive reduced mass [$1/m({\rm e})+1/m({\rm h})$]. In a classical picture, the electron-hole pairs at $\Gamma$ are expected to form excitons that jointly orbit around a common center which does not lie between the two particles\,\cite{Lin2021}.

Because the lowest conduction band is relatively isolated from the other conduction bands, we can define its band width $w_{\rm CB}$. We display $w_{\rm CB}$ as a function of the nanoribbon widths $W$ in Fig.\,\ref{qTP-H}(d). The smallest $w_{\rm CB}$ of 81\,meV is observed for $W=2$. The band width increases monotonically with the width of the nanoribbon $W$, as the lowest conduction band becomes more dispersive, approaching that of the monolayer, 95\,meV.

We next study the band gaps of qTP-H nanoribbons. As summarized in Fig.\,\ref{qTP-H}(b), there are four possible electronic transitions, depending on the point of the Brillouin zone at which they occur: the two transitions involving direct band gaps at $\Gamma$ and Y, denoted as $E_{\rm g}^{\Gamma}$ and $E_{\rm g}^{\rm Y}$ respectively; the transition involving indirect band gaps from h$_{\rm Y}$ to e$_{\Gamma}$, denoted as $E_{\rm g}^{\rm Y-\Gamma}$; and the transition involving indirect band gaps from h$_{\Gamma}$ to e$_{\rm Y}$, denoted as $E_{\rm g}^{\rm \Gamma - Y}$.  The smallest band gap is $E_{\rm g}^{\rm \Gamma - Y}$ for all $W$, which converges to the monolayer gap of 1.06\,eV for $W>9$. The $E_{\rm g}^{\rm \Gamma - Y}$ of 1.29\,eV for $W=2$ is comparable to that obtained in previous calculations of the 1D fullerene chain with $W=1$\,\cite{Jones2023}. Previous computational studies have shown that for both 1D chain and various 2D networks of C$_{60}$, the band gap difference between unscreened hybrid functional (HF) and DFT is around 1.23\,eV\,\cite{Jones2023}. Therefore, we can shift the band gaps rigidly with this correction to estimate the HF band gaps of quasi-1D nanoribbons accurately, which provide agreeable results\,\cite{Peng2022c,Jones2023,Shearsby2025} with the measured band gaps\,\cite{Hou2022,Meirzadeh2023,Wang2023}. The direct band gaps at $\Gamma$ also converges to the monolayer gap with increased $W$. This is unsurprising, as their corresponding band edges are in similar positions. The band gaps $E_{\rm g}^{\rm  Y - \Gamma}$ and $E_{\rm g}^{\rm Y}$ for the quasi-1D nanoribbons do not converge at the monolayer gaps. This is because the VBM at Y for the nanoribbons becomes higher than the monolayer VBM at Y.

\begin{figure*}
\centering
\includegraphics[width=\linewidth]{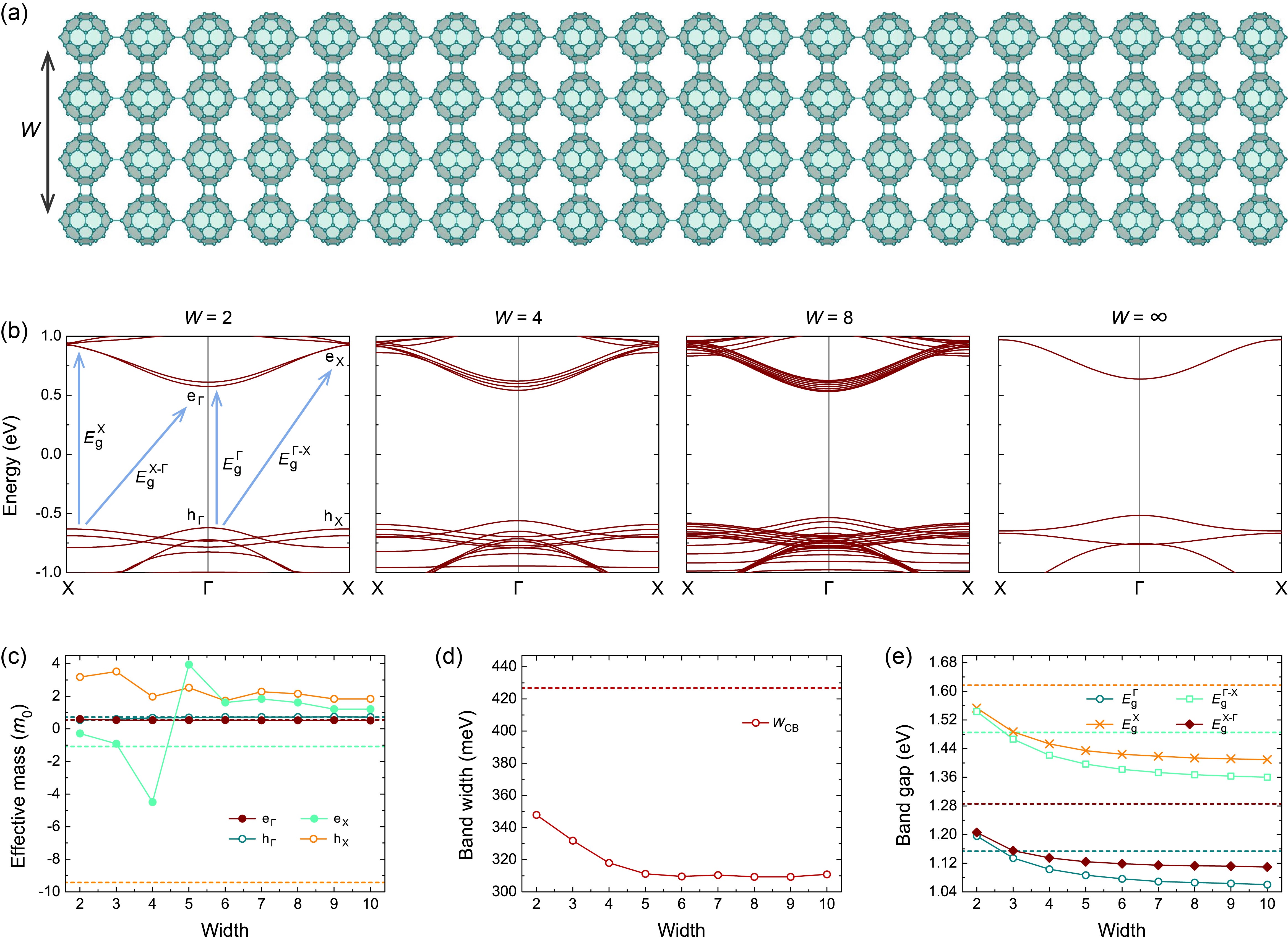}
\caption{
(a) Crystal structures of qTP-V nanoribbons with a representative width  $W$ of 4. (b) Band structures, (c) effective masses, (d) band widths, and (e) band gaps of qTP-V nanoribbons as a function of $W$.
}
\label{qTP-V} 
\end{figure*}

\vspace{1mm}

\subsection{qTP-V} 
Different from qTP-H, the carbon cages in the qTP-V nanoribbons are connected by vertical [2\,+\,2] cycloaddition bonds along the nanoribbon direction, while the cages perpendicular to the nanoribbons are linked by the horizontal cycloaddition bonds, as shown in Fig.\,\ref{qTP-V}(a). The qTP-V nanoribbons with $W=4$ have the same space group with qTP-H but a slightly longer lattice constant of 9.13\,\AA\ and a slightly smaller width of 33.90\,\AA, suggesting a structural asymmetry between qTP-V and qTP-H nanoribbons.

The band structures of qTP-V nanoribbons exhibit distinct behaviors compared to qTP-H nanoribbons. We find direct band gap features for all $W$ in Fig.\,\ref{qTP-V}(b), mainly because the e$_{\Gamma}$ state has much lower energy than e$_{\rm X}$. With increased $W$, the electronic structure have more replicas of the bands, while the direct band gap remains stable. For e$_{\rm X}$ with $W=4$, extra bands from higher conduction states become lower in energy, leading to a sudden increase of the corresponding effective masses.

Fig.\,\ref{qTP-V}(c) shows the evolution of effective masses as a function of $W$. The negative $m({\rm e}_{\rm Y})$ changes its sign abruptly when $W$ increases from 4 to 5 owing to the lowering of higher conduction bands as expected. This leads to a different $m({\rm e}_{\rm Y})$ from the monolayer, where the lowest conduction band is isolated from higher bands. Similarly, the h$_{\rm X}$ for finite $W$ with positive curvature is contributed by lower valence bands as well, but the highest valence band is relatively isolated for $W = \infty$ with small negative curvature around X. For the band edges at $\Gamma$, the effective masses are nearly a constant from $W=2$ to $W = \infty$, as the VBM and CBM at $\Gamma$ of the qTP-V nanoribbons are rigid shifts of the replicas.

Despite that all the bands in qTP-V nanoribbons cross with other bands, we can still choose the relatively isolated lowest conduction band and determine its band width. As shown in Fig.\,\ref{qTP-V}(d), the $w_{\rm CB}$ decreases because of the lowering of the conduction bands at X from the crossed higher conduction bands, leading to much smaller band width of $\sim 310$\,meV compared to that observed in the monolayer, 427\,meV.

All the band gaps of qTP-V nanoribbons reported in Fig.\,\ref{qTP-V}(e) deviate significantly from the monolayer case. The smallest band gap results from the direct transition at the $\Gamma$ point. At $W=2$, the direct $E_{\rm g}^{\Gamma}$ is only slightly smaller than the indirect $E_{\rm g}^{\rm X-\Gamma}$. However, their difference becomes larger when $W$ is increased, leading to more distinct direct band gap features. On the other hand, the direct band gap at X is always the largest, and the difference between the direct $E_{\rm g}^{\Gamma}$ and the second largest indirect $E_{\rm g}^{\rm \Gamma - X}$ also increases for larger $W$.

\begin{figure*}
\centering
\includegraphics[width=\linewidth]{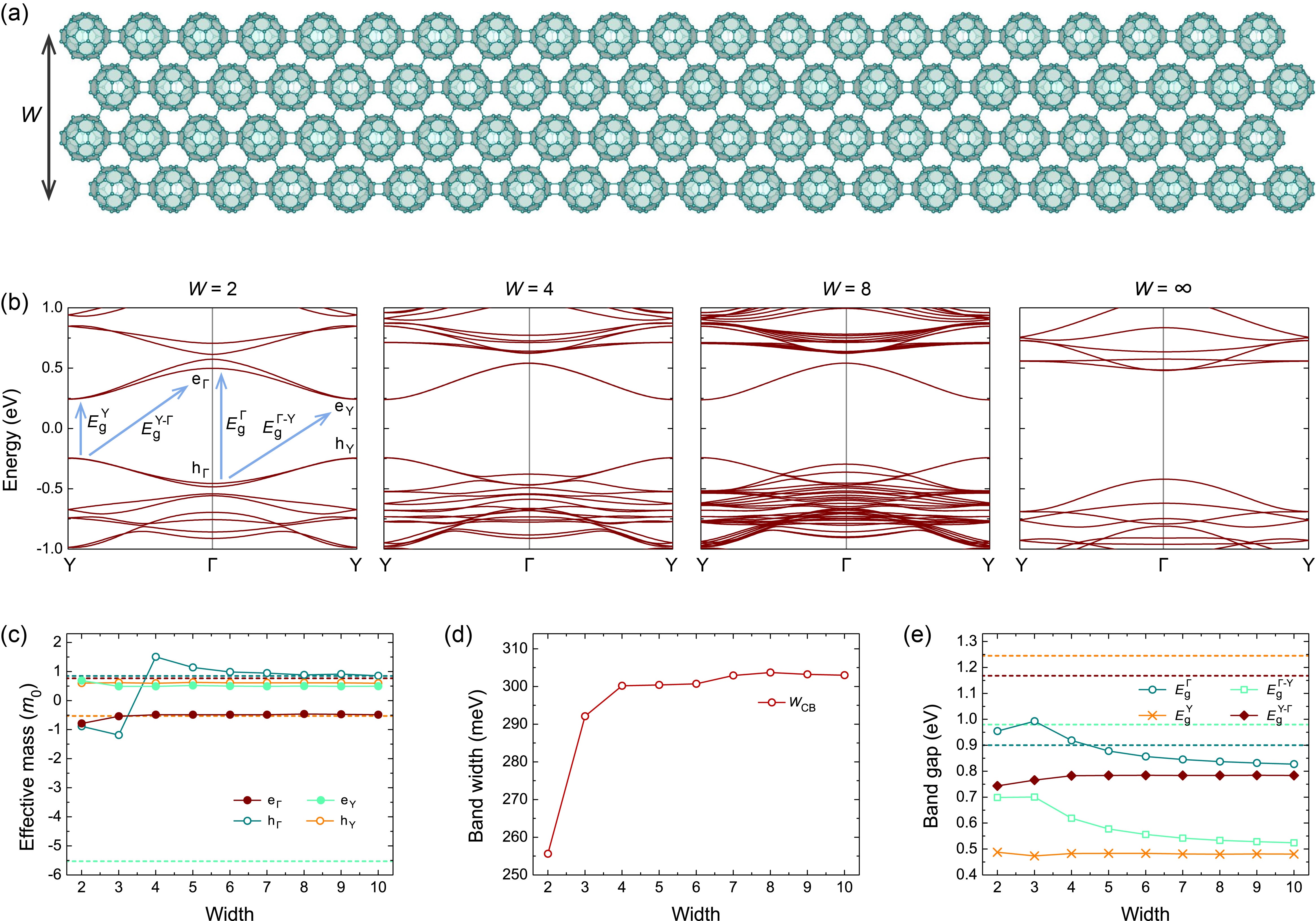}
\caption{
(a) Crystal structures of qHP-AC nanoribbons with a representative width  $W$ of 4. (b) Band structures, (c) effective masses, (d) band widths, and (e) band gaps of qHP-AC nanoribbons as a function of $W$.
}
\label{qHP-AC} 
\end{figure*}

\vspace{1mm}

\subsection{qHP-AC} 
Monolayer qHP networks exhibit structural features differing from qTP. The C$_{60}$ molecules are arranged in a space-efficient manner, as shown in Fig.\,\ref{crystals}(d). We first focus on qHP-AC nanoribbons with the armchair-like edges, the crystal structure of which is shown in Fig.\,\ref{qHP-AC}(a). The molecular cages along the nanoribbons are connected by nearly-planar [2\,+\,2] cycloaddition bonds, while the cages across the nanoribbons are linked by intermolecular C$-$C single bonds. The space group of qHP-AC nanoribbons is $P222_1$ (No.\,17) with a lattice constant of 9.15\,\AA. A width of 30.75\,\AA\ for $W=4$. The qHP-AC width is shorter than those in qTP nanoribbons owing to the closely-packed qHP structures. 

The band structures of qHP-AC nanoribbons exhibit extra in-gap states compared to qHP monolayers in Fig.\,\ref{qHP-AC}(b). These in-gap states are almost purely contributed by the molecules on the two edges of the nanoribbons. While the lower valence bands and higher conduction bands show band replicas with the number of replicas proportional to $W$, the number of in-gap states is fixed for all $W$. For $W=2$, there are two nearly degenerate in-gap valence states and two in-gap conduction states in an energy window between $-0.5$ and 0.5\,eV. With increasing $W$, the two in-gap valence/conduction states become completely degenerate, as the interactions between the edge states  is reduced when the two edges are separated.

We then consider the evolution of effective masses with increasing $W$. Because the CBM at both $\Gamma$ and Y for qHP-AC nanoribbons are contributed purely by the edge states, their corresponding effective masses converge quickly to a constant value when $W>2$. The same conclusion holds for the VBM at Y. However, the effective masses $m({\rm e}_{\Gamma})$, $m({\rm e}_{\rm Y})$, and $m({\rm h}_{\rm Y})$ of the nanoribbons are completely different from those of the qHP monolayer. This is expected as these effective masses are from the edge states instead of the monolayer states. For h$_{\Gamma}$, the monolayer states at $\Gamma$ become higher than the edge states when $W>3$. This leads to an abrupt change of signs of the effective mass $m({\rm h}_{\Gamma})$ from $W=3$ to $W=4$. Interestingly, the effective masses $m({\rm h}_{\Gamma})$ of the qHP-AC nanoribbons start to converge to that of the monolayers when $W>6$, as the VBM at $\Gamma$ shows similar curvatures.

For qHP-AC nanoribbons, the edge states of the lowest conduction band are quite isolated from the monolayer bands. The band width $w_{\rm CB}$ of 255\,meV for $W=2$ increases with larger $W$, leading to a converged band width $w_{\rm CB}$ of 303\,meV for $W>6$ [Fig.\,\ref{qHP-AC}(d)]. This also leads to a nearly constant $w_{\rm CB}$ similar to the fixed $m({\rm e}_{\Gamma})$ and $m({\rm e}_{\rm Y})$.

Similar to the effective masses and the band widths, the band gaps of qHP-AC nanoribbons also converge to those of the edge states, as shown in Fig.\,\ref{qHP-AC}(e). The direct band gap at $\Gamma$ is the smallest band gap for qHP monolayers, whereas the direct band gap at Y is the smallest band gap for qHP-AC nanoribbons. The presence of edge states leads to a band gap difference of 420\,meV between qHP monolayers and qHP-AC nanoribbons, which might explain the difference in the measured electronic band gaps of $1.60-2.05$\,eV\,\cite{Hou2022,Meirzadeh2023,Wang2023} and optical band gaps of $1.10-1.55$\,eV\,\cite{Hou2022,Zhang2025} due to the finite size of the samples. As the energy of h$_{\Gamma}$ becomes higher with increased $W$, the band gap $E_{\rm g}^{\rm \Gamma - Y}$ decreases, and the difference between $E_{\rm g}^{\rm \Gamma - Y}$ and $E_{\rm g}^{\rm Y}$ reduces.

\begin{figure*}
\centering
\includegraphics[width=\linewidth]{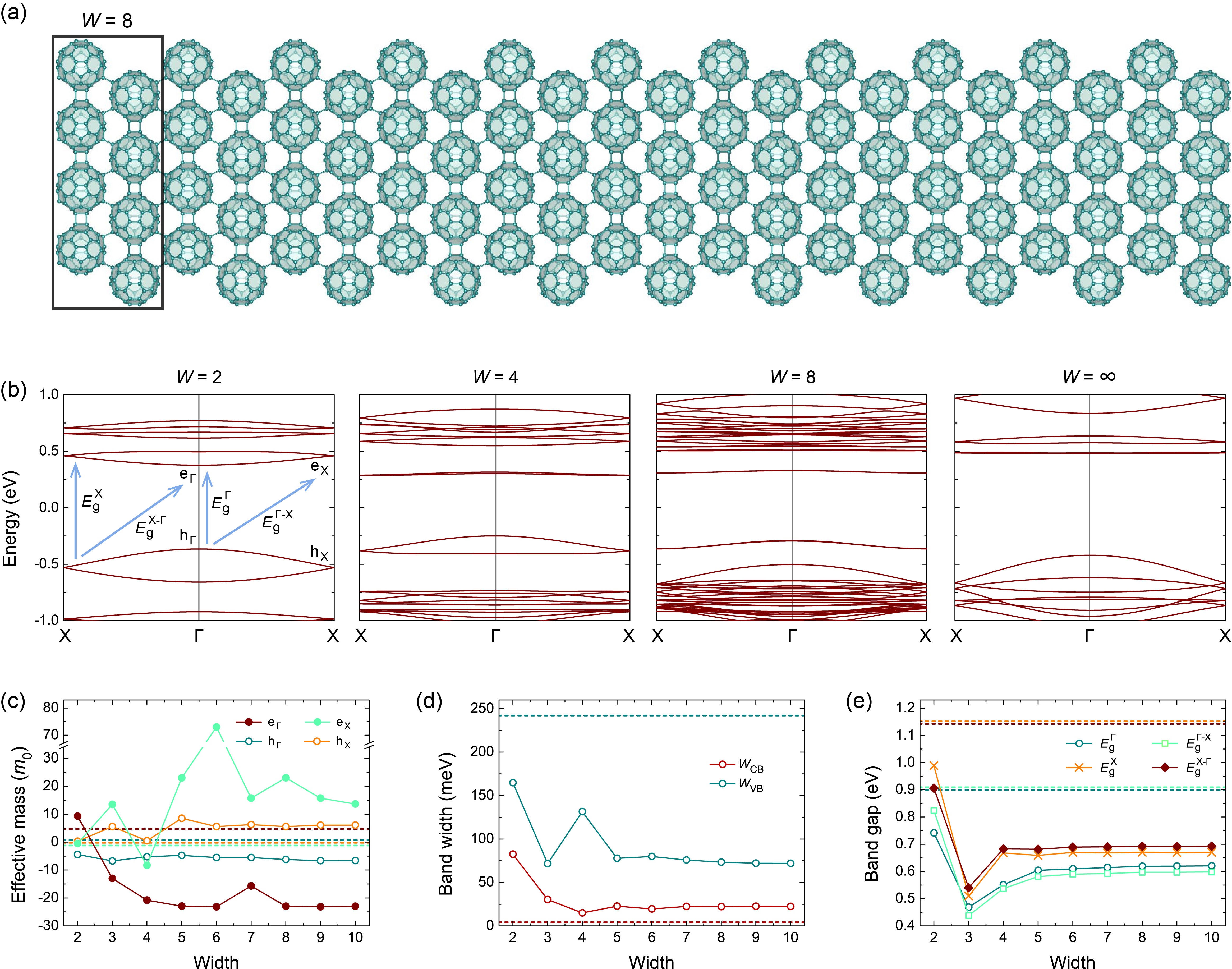}
\caption{
(a) Crystal structures of qHP-ZZ nanoribbons with a representative width $W$ of 8. (b) Band structures, (c) effective masses, (d) band widths, and (e) band gaps of qHP-ZZ nanoribbons as a function of $W$.{\rm CB}
}
\label{qHP-ZZ} 
\end{figure*}

\vspace{1mm}

\subsection{qHP-ZZ} 
Fig.\,\ref{qHP-ZZ}(a) shows the crystal structures of qHP-ZZ nanoribbons. The zigzag edge is made of the triangular edges of neighboring C$_{60}$ connected by the diagonal single bonds. The nanoribbon width $W$ is defined as the number of fullerene molecules in the primitive unit cell of the nanoribbon, as demonstrated by $W=8$ in Fig.\,\ref{qHP-ZZ}(a). The qHP-ZZ nanoribbons with odd $W$ have a space group of $P2/m$ (No.\,10), with inversion symmetry, as well as $C_2$ and mirror symmetry with respect to $y$ axis. On the other hand, nanoribbons with even $W$ have a space group of $Pma2$ (No.\,28) with $C_2$ rotational symmetry along $z$ and glide mirror symmetry with respect to both $x$ and $y$ axis. 
The lattice constant for $W=8$ is 15.85\,\AA, and the width of 38.96\,\AA\ is larger than the qTP-H, qTP-V, and qHP-AC nanoribbons for $W=4$ because of the additional fullerene cage.

Similar to the qHP-AC nanoribbons, the band structures of qHP-ZZ nanoribbons also have in-gap edge states, as shown in Fig.\,\ref{qHP-ZZ}(b). There are two in-gap valence bands and two in-gap conduction bands. With increasing $W$, the two conduction bands become nearly degenerate for $W>4$, while the two valence bands become degenerate for $W>7$. For $W>5$, the band structures start to be independent of the oddness and evenness of $W$, with the monolayer features for $W=\infty$ clearly visible, as shown by the $W=8$ panel in Fig.\,\ref{qHP-ZZ}(b).

Because of the in-gap states, the effective masses of the nanoribbons show distinct features from their monolayer counterpart, as shown in Fig.\,\ref{qHP-ZZ}(c). Notably, the in-gap conduction bands are much less dispersive than the other bands, leading to large positive effective mass for e$_{\rm X}$ with $m({\rm e}_{\rm X})> 10\,m_0$ for $W>4$, as well as large negative effective mass for e$_{\Gamma}$ with $m({\rm e}_{\Gamma})<-20\,m_0$ for $W>7$. In particular, $m({\rm e}_{\rm X})$ becomes higher than 70\,$m_0$ for $W=6$, yielding a nearly-flat band.

The edge states of the qHP-ZZ nanoribbons become less dispersive with increasing $W$, as demonstrated by the band width $w_{\rm CB}$ and $w_{\rm VB}$ in Fig.\,\ref{qHP-ZZ}(d). The conduction band width converges to 72\,meV for $W>8$, while the valence band width converges to 22\,meV for $W>6$. Interestingly, the $w_{\rm CB}$ of 5\,meV for qHP monolayers along $\Gamma$--X is even smaller than that of the qHP-ZZ nanoribbons, while the monolayer $w_{\rm VB}$ of 242\,meV is much larger than that of the nanoribbons.

The evolution of the band gaps with increasing $W$ in Fig.\,\ref{qHP-ZZ}(e) shows the change from a direct band gap at $\Gamma$ ($E_{\rm g}^{\Gamma}$) to an indirect band gap ($E_{\rm g}^{\rm \Gamma - X}$) when $W$ increases from 2 to 3. Further increasing $W$ results in the same indirect $E_{\rm g}^{\rm \Gamma - X}$ as the smallest band gap, while the band gap differences among $E_{\rm g}^{\Gamma}$, $E_{\rm g}^{\rm X}$, $E_{\rm g}^{\rm \Gamma - X}$, and $E_{\rm g}^{\rm X -\Gamma}$ are within 94\,meV. Owing to the presence of the in-gap edge states, the band gap difference between the smallest monolayer gap $E_{\rm g}^{\rm \Gamma}$ and the converged, smallest nanoribbon gap $E_{\rm g}^{\rm \Gamma - X}$ is 303\,meV.

\section{Conclusions}

\noindent In summary, we systematically investigate the electronic structure of fullerene nanoribbons derived from two monolayer phases for different crystalline directions with varied widths on the basis of first-principles calculations. For qTP fullerene networks, fabricating nanoribbons along the vertical or horizontal intermolecular [2\,+\,2] cycloaddition bonds leads to distinct electronic properties, with direct band gaps for qTP-V nanoribbons and indirect band gaps for qTP-H nanoribbons respectively. For qHP nanoribbons, there are extra in-gap edge states for both conduction and valence bands, and such edge states result in direct band gaps for qHP-AC nanoribbons but indirect band gaps for qHP-ZZ nanoribbons with $W>2$. Our work reveals a rich variety of electronic properties emerging in fullerene nanoribbons depending on the details of their crystal structures, possibly laying the foundation for the design of scalable fullerene-based nanodevices.

\section{Acknowledgements}

\noindent B.P. acknowledges support from Magdalene College Cambridge for a Nevile Research Fellowship. The calculations were performed using resources provided by the Cambridge Service for Data Driven Discovery (CSD3) operated by the University of Cambridge Research Computing Service (\url{www.csd3.cam.ac.uk}), provided by Dell EMC and Intel using Tier-2 funding from the Engineering and Physical Sciences Research Council (capital grant EP/T022159/1), and DiRAC funding from the Science and Technology Facilities Council (\url{http://www.dirac.ac.uk} ), as well as with computational support from the UK Materials and Molecular Modelling Hub, which is partially funded by EPSRC (EP/T022213/1, EP/W032260/1 and EP/P020194/1), for which access was obtained via the UKCP consortium and funded by EPSRC grant ref EP/P022561/1.


%

\end{document}